\documentclass[12pt]{article}
 
\usepackage[margin=0.75in]{geometry} 
\usepackage{amsmath,amsthm,amssymb,color}
\usepackage{graphicx}
\usepackage{mathtools}
\usepackage[section]{placeins}
\usepackage{caption}
\usepackage{subcaption}
\usepackage{cleveref}
\usepackage{bbm}
\usepackage{hyperref}

\newcommand{\E}{\mathbb{E}}

\begin{document}
 
 
\title{Detecting Community Structures in Hi-C Genomic Data}
\author{Irineo Cabreros\footnote{Program in Applied and Computational Mathematics, Princeton University. Email: cabreros@princeton.edu}  \and Emmanuel Abbe\footnote{Program in Applied and Computational Mathematics, and Department of Electrical Engineering, Princeton University. Email: eabbe@princeton.edu} \and Aristotelis Tsirigos \footnote{Department of Pathology, NYU Langone Medical Center. Email: Aristotelis.Tsirigos@nyumc.org}} 

\date{}
 
\maketitle

\abstract{Community detection (CD) algorithms are applied to Hi-C data to discover new communities of loci in the 3D conformation of human and mouse DNA. We find that CD has some distinct advantages over pre-existing methods: (1) it is capable of finding a variable number of communities, (2) it can detect communities of DNA loci either adjacent or distant in the 1D sequence, and (3) it allows us to obtain a principled value of $k$, the number of communities present. Forcing $k=2$, our method recovers earlier findings of Lieberman-Aiden, et al. (2009), but letting $k$ be a parameter, our method obtains as optimal value $k^* = 6$, discovering new candidate communities. In addition to discovering large communities that partition entire chromosomes, we also show that CD can detect small-scale topologically associating domains (TADs) such as those found in Dixon, et al. (2012). CD thus provides a natural and flexible statistical framework for understanding the folding structure of DNA at multiple scales in Hi-C data.}

\section{Introduction}
The 3D conformation of DNA plays a vital role in gene regulation through many varied mechanisms \cite{dekker}. For example, genes may be silenced or expressed depending on whether they are positioned in tightly packed (closed) or loosely packed (open) regions since exposure to transcriptional machinery influences transcriptional rates. As another example, several studies report examples of enhancers positioned up to a megabase away from their target genes nevertheless capable of modulating transcription rates \cite{lettice, Sotelo, nobrega}. This scenario is thought to be explained by chromatin looping events that position distant elements in the 1D sequence in close proximity in the full 3D conformation \cite{kadauke}. Features of the 3D conformation of DNA have been shown to be highly conserved evolutionarily and it has even been demonstrated that some chromatin structures are heritable independent of DNA sequence itself \cite{felsenfeld}. Furthermore, the 3D conformation of DNA is of direct clinical importance as defects in chromatin structure have been identified with premature aging, limb malformation, and cancer \cite{scaffidi,lupianez,meaburn}.

Though it is currently not possible to directly observe the 3D folding structure of DNA, this work shows that community detection techniques can be applied to Hi-C data to find clusters of loci\footnote{By ``loci," we refer segments of base pairs that are adjacent in the 1D DNA sequence. Different Hi-C technologies have different scales of resolution ranging from $\sim$ 1 Mbp to $\sim$ 10 kbp \cite{erez, dixon}.} that are proximal in the 3D conformation of DNA. Hi-C is a recently developed method that indirectly captures information about the 3D conformation of DNA by observing genome-wide chromatin interactions using a technique of spatially constrained ligation in conjunction with massively parallel sequencing \cite{erez}. The output of this process is a square matrix $X$ whose $(i,j)$ element is the number of ligation events observed between segment $i$ and segment $j$. This matrix can be interpreted as a weighted, undirected adjacency matrix of a graph whose nodes represent individual DNA segments. Previous techniques to detect cluster the Hi-C connection matrix are catered to finding either (1) two large compartments \cite{erez} (which typically correspond to open and closed ``sectors" of DNA) each of which are fragmented throughout an entire chromosome or (2) an indeterminate number of clusters of adjacent DNA called ``topologically associating domains" (TADs) \cite{dixon, filippova}\footnote{We note that while this work was in progress, an independent parallel work suggested applying several standard clustering methods (HMM, k-means, and hierarchical clustering) which are also capable of finding a variable number of communities \cite{erez2}. Though the precise connection between these findings and our own is not fully examined, we briefly note some prominent differences here. First, clustering methods are applied to a matrix $C$ which is a highly modified version of the original contact matrix. By contrast, our method applies directly to the thresholded connection matrix. Second, our method for selecting $k^*=6$ is quite different from theirs, which relies both on the Akaike Information Content to suggest a range of viable $k$ along with visual inspection. Finally, our method directly fits a statistical network model that closely mirrors the physical network of interconnected DNA loci underlying Hi-C data; individual loci belong to (potentially multiple) communities and probabilistically undergo ligation events (i.e. form an edge) with members of their communities during the Hi-C procedure.}. We propose the use of community detection techniques as a much more flexible approach to finding a variable number communities by fitting statistical models of mixed-membership networks. 

Detecting communities (or clusters) in graphs is a fundamental and long studied problem in computer science and machine learning. The techniques apply to a large variety of complex networks, such as social or biological networks, or to data sets engineered as networks via graphs of similarity. Extracting communities in networks has been used in particular to find like-minded people in social networks \cite{newman-girvan,social1}, to perform recommendations \cite{amazon}, to segment or classify images \cite{image1,image2}, to detect protein complexes \cite{ppi2,marcotte}, to find genetically related sub-populations \cite{genetics,gene-survey}, to discover new tumor subclasses \cite{tumor}, among other applications \cite{newman1,fortunato,comparatif}.

This paper introduces the use of community detection methods to the network obtained from the Hi-C contact matrices, extracting communities by fitting probabilistic models such as mixed-membership stochastic block models \cite{prem} and Poisson models \cite{ball}. We refer to Appendix \ref{algos} for a more detailed description of the models. In short, these are extensions of the standard stochastic block model, introduced in \cite{holland}, which assigns to each vertex in the network a community variable that affects the edge probabilities. 

The data analyzed in this work are from Lieberman-Aiden, et al. \cite{erez} and Dixon, et al. \cite{dixon}. We focus primarily on data from the former and study in particular chromosome 14 from the human cell line GM06990. From the latter, we study a segment of chromosome 6 from a mouse embryonic stem cell line. Two algorithms are applied to these data and show results consistent with previous findings while proposing novel communities at previously unexamined scales. The first algorithm (developed by P. Gopalan and D.M. Blei) uses stochastic variational inference on a mixed-membership stochastic block (MMSB) model \cite{prem} and the second (developed by B. Ball, B. Karrer, and M. E. J. Newman) is an expectation-maximization algorithm applied to a Poisson model for network generation \cite{ball}. Both algorithms are described in further detail in Appendix \ref{algos}.

\section{Overview of our contribution}

\begin{figure}[t] 
   \centering
   \includegraphics[width=4in]{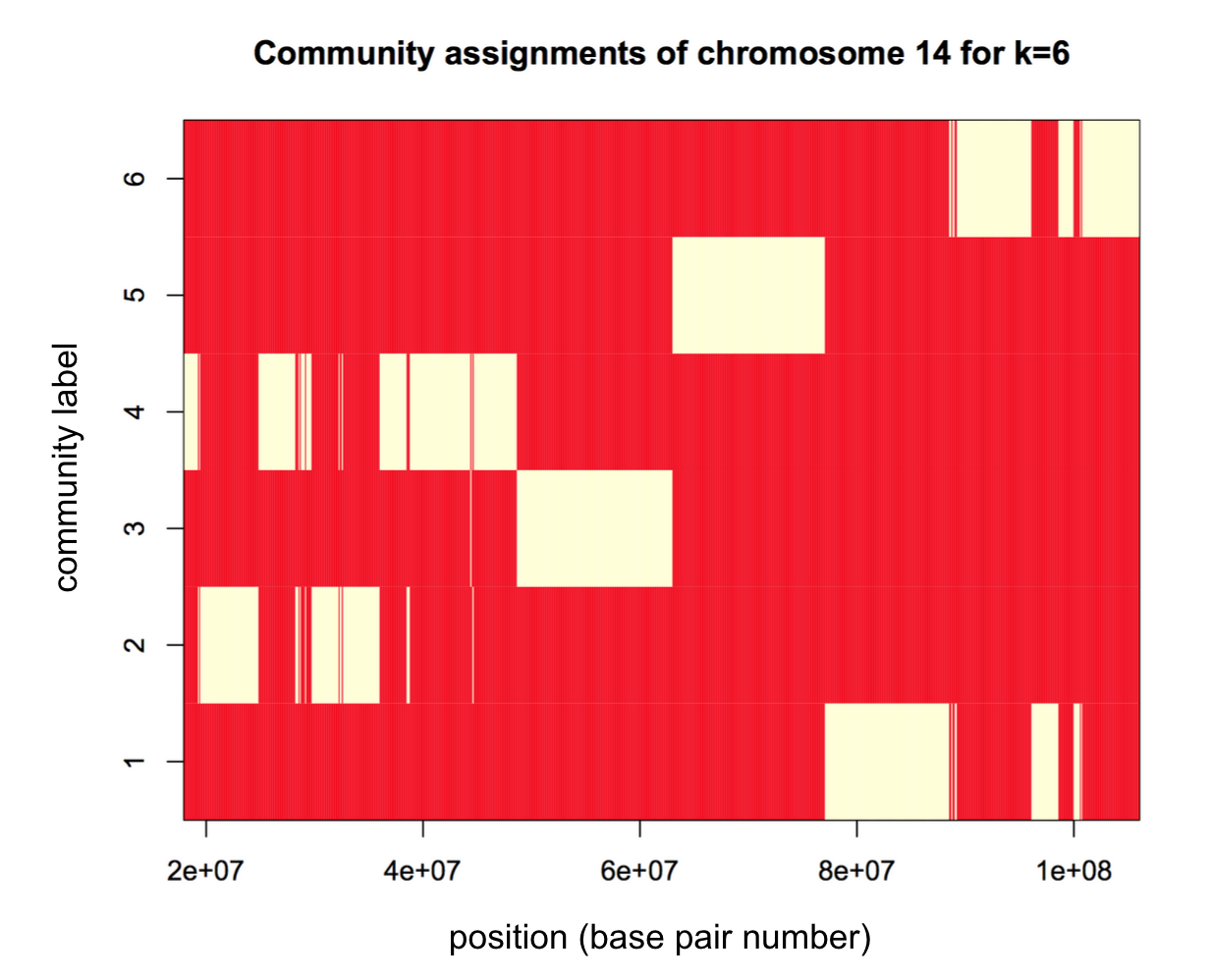} 
   \caption{Community assignment matrix for $k=k^*=6$. Each row of this matrix represents a community and the columns index genome position. An element $(i,j)$ of this matrix is beige if the CD algorithm identifies locus $j$ as having maximal membership to community $i$ and is red otherwise.}
   \label{k6}
\end{figure}

In this work, we introduce the application of community detection techniques to Hi-C data to discover novel chromatin structures in the 3D conformation of DNA. Below we outline the program of this paper: 
\begin{itemize}
\item We first show that CD offers a unified method that is capable of reproducing the results of two previous techniques tailored to identifying particular types of communities. 
\begin{itemize}
\item Section \ref{nonlocal} shows that CD finds two large, genome-wide compartments of open and closed chromatin first identified in the work of Lieberman-Aiden, et al. 

\item Section \ref{local} we show that CD is capable of discovering topologically associating domains (TADs): highly localized communities of a much smaller scale than compartments. An underlying message of both \ref{nonlocal} and \ref{local} is that the results appear to be robust to the particular CD algorithm used; both the MMSB algorithm and Poisson algorithm converge to either identical or very similar results. 
\end{itemize}

\item In section \ref{new}, we exploit the flexibility of CD to discover new community structures. In particular, we introduce two different approaches to address the model selection problem of choosing $k$ and show that, in conjunction, they suggest the distinguished value $k^*=6$ for chromosome 14 of cell line GM06990 (a chromosome focused on in the paper by Lieberman-Aiden, et al.). Fig. \ref{k6} shows the community assignment matrix for this chromosome for $k=6$ and we note that it contains both locally interacting (communities 2 and 4) as well as non-locally interacting (communities 1, 3, 5, and 6) clusters of DNA loci.
\end{itemize}

\section{Results}

\subsection{CD algorithms successfully identify non-local compartments}
\label{nonlocal}
The analysis of Lieberman-Aiden et al. suggests that the DNA within a single chromosome is divided into two separate ``compartments" \cite{erez}. In order to find these compartments, the authors first normalized the observed Hi-C contact matrix (by dividing the counts of each matrix element $(i,j)$ by the chromosome-wide expected number of counts at a distance $|i-j|$), then computed the Pearson correlation matrix of the resultant matrix, and finally performed principal component analysis (PCA) on this matrix. The first principle component of the correlation matrix was then used to partition the chromosome into two compartments: the first compartment defined by the loci with a positive coefficient in the first principle component and the second compartment defined by the loci with a negative coefficient in the first principle component. The resultant compartments displayed several consistent biological characteristics. The first compartment corresponds to more gene-rich portions of the chromosome and higher expression levels. Consistent with the interpretation that DNA within the first compartment is more accessible (i.e. has a more ``open" conformation), DNA within the first compartment was also found to have a significantly lower interaction frequency than DNA within the second compartment.

\begin{figure}[h] 
   \centering
   \includegraphics[width=7in]{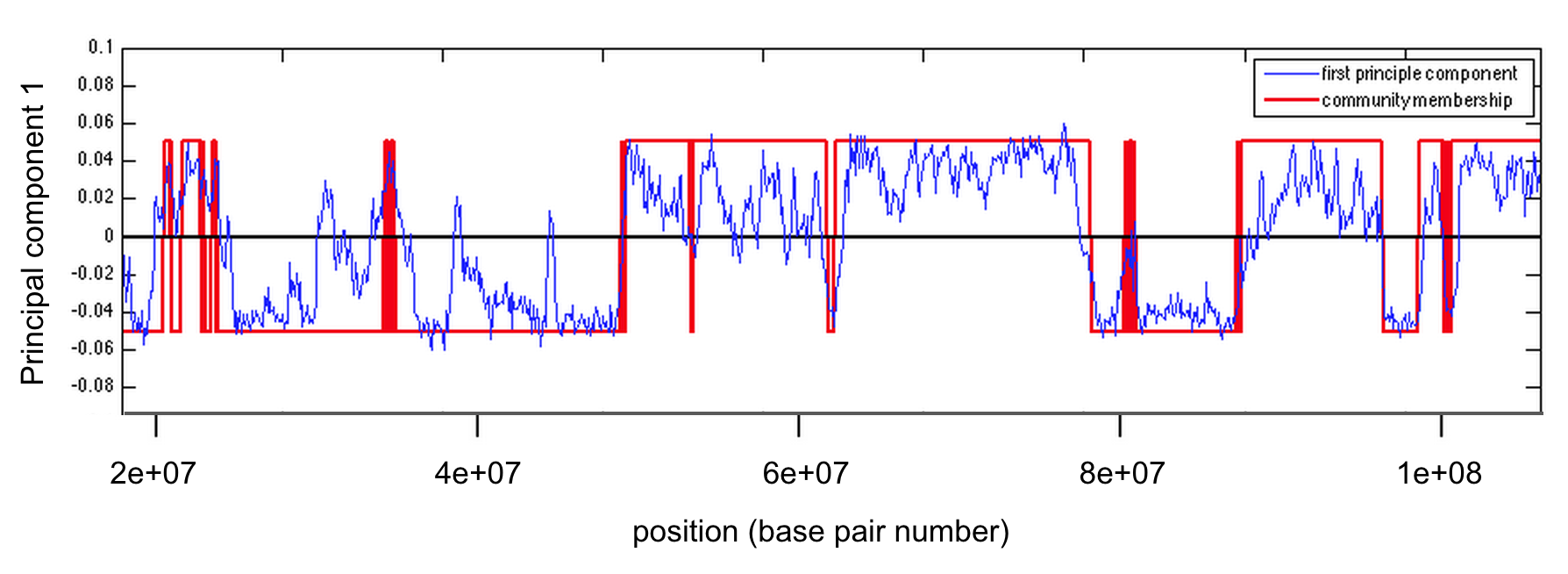} 
   \caption{Comparison between results of Lieberman-Aiden, et al. and CD algorithms. Red line corresponds to community membership: a value of 0.5 corresponds to membership in community one and a value of -0.5 corresponds to membership in community two. The blue line corresponds to the values of the first principle component of the Pearson correlation matrix. In Lieberman-Aiden, et al. they identify loci with positive/negative values of the first principle component as belonging to ``compartment" A/B. It is apparent that the community assignments of our algorithms closely follow the contour of the first principle component.}
   \label{princomp}
\end{figure}

Although the analysis of Lieberman-Aiden et al. was very successful in identifying a biologically relevant partition of DNA, it has some disadvantages to the CD algorithms discussed in this paper. First, the underlying model of the CD algorithms lends a more straightforward statistical interpretation than the procedure used by Liebermann-Aiden et al.; where the former seeks MLE estimates of parameters of a well-defined statistical network model that closely mirrors the physical scenario, it is not obvious what assumptions are required for the modified PCA procedure to yield a valid partitioning of the chromosome. While the steps of the procedure utilized by Lieberman-Aiden et al. are motivated by their ability to create clear ``plaid" patterns in the connection matrix (see Fig. \ref{princomp2}), it is not obvious a priori why this should be a desirable goal in the first place. Second, it is unclear how to generalize the PCA procedure to find more than two communities whereas both CD algorithms can be set to find an arbitrary number of communities. Fig.\ref{princomp} compares the result of the CD algorithm with the first principle component of chromosome 14 (human DNA). Since both algorithms fit mixed membership models, we decide to simplify the presentation by identifying locus $i$ with a single community $a = \underset{i}{\text{argmax }} \theta_i(k)$ rather than reporting the full mixed-membership vector for locus $i$.  Though we do not exploit the mixed-membership nature of our CD algorithms, it may be useful in future studies. A recent work shows evidence for precisely choreographed chromatin reorganizations in time (the so-called ``4D" structure of DNA) \cite{chen}. It may be the case that regions of loci that are highly admixed (i.e. the vector $\theta_i$ has more than one large component) will be good candidates as chromatin structures that reposition themselves dynamically. 

In this particular case, both algorithms resulted in identical plots. As can be seen, the output of the CD algorithm closely follows the contour of the first principle component, discovering a very close partition to that of the PCA procedure. While the PCA procedure requires as input the Pearson correlation matrix of a normalized Hi-C connection matrix, the CD algorithms take as input a binarized version of the original matrix $X^*_{ij} = \mathbbm{1}(X_{ij} > 0)$.This binarization is used because the current implementation of the MMSB algorithm only accepts binary adjacency matrices as input (see section \ref{MMSB}). It should be noted that this is not a fundamental limitation of CD algorithms and that several other existing CD algorithms, including the Poisson model that is discussed in this paper, allow for weighted adjacency matrices. We do not use the weighted adjacency matrix for the Poisson algorithm in this study for simplicity of comparison with the MMSB algorithm. In future work, it would be interesting to exploit these intensities as well. This could be done using the Poisson model, or with a tailored model that allows to integrate properly the intensity variations occurring in the Hi-C contact matrices. We believe that reworking these analyses incorporating intensity information matrices will be important for future investigations. It is also interesting that the communities found by the CD algorithms as well as PCA are non-local: both communities include large DNA segments from disparate regions of the chromosome. As can be seen in supplementary Fig. \ref{princomp2} both the principle component vector and the community memberships clearly demarcate the plaid pattern of the correlation matrix. 

\subsection{CD algorithms successfully identify local TADs}
\label{local}
Unlike the large, non-local compartments studied above, Dixon et al. study small ($\sim1$Mb) and local (i.e. adjacent in the linear DNA structure) communities called topologically associating domains (TADs). TADs are of biological interest because they are conserved over several cell types as well as several species suggesting that their structure is important to basic mammalian cell function \cite{dixon}. Additionally, several disorders such as limb malformation have been associated with structural deformities in chromatin at the scale of TADS \cite{lupianez}. The method by which Dixon et al. discover topological domains is to take advantage of a bias in the directionality of interactions at the boundary of TADs. At the downstream terminus of an interacting unit, most interactions will be with upstream portions of DNA. Conversely at the upstream terminus of an interacting unit, most interactions will be with downstream portions of DNA. Dixon et al. define a directionality index (DI) which captures the directionality of links at any particular loci and then use a hidden Markov model to infer boundary locations in which the DI quickly changes. 

While the method described above is successful at identifying TADs, it has one significant disadvantage to the methods of CD algorithms.  Since this method identifies boundaries and not communities, it can only determine local structures (i.e. it will never identify portions of the genome that are distant from each other in the linear structure of the DNA as belonging to the same community). By contrast, CD algorithms (as was displayed in the previous section), naturally detect non-local community structures. 

\begin{figure}[h] 
   \centering
   \includegraphics[width=7in]{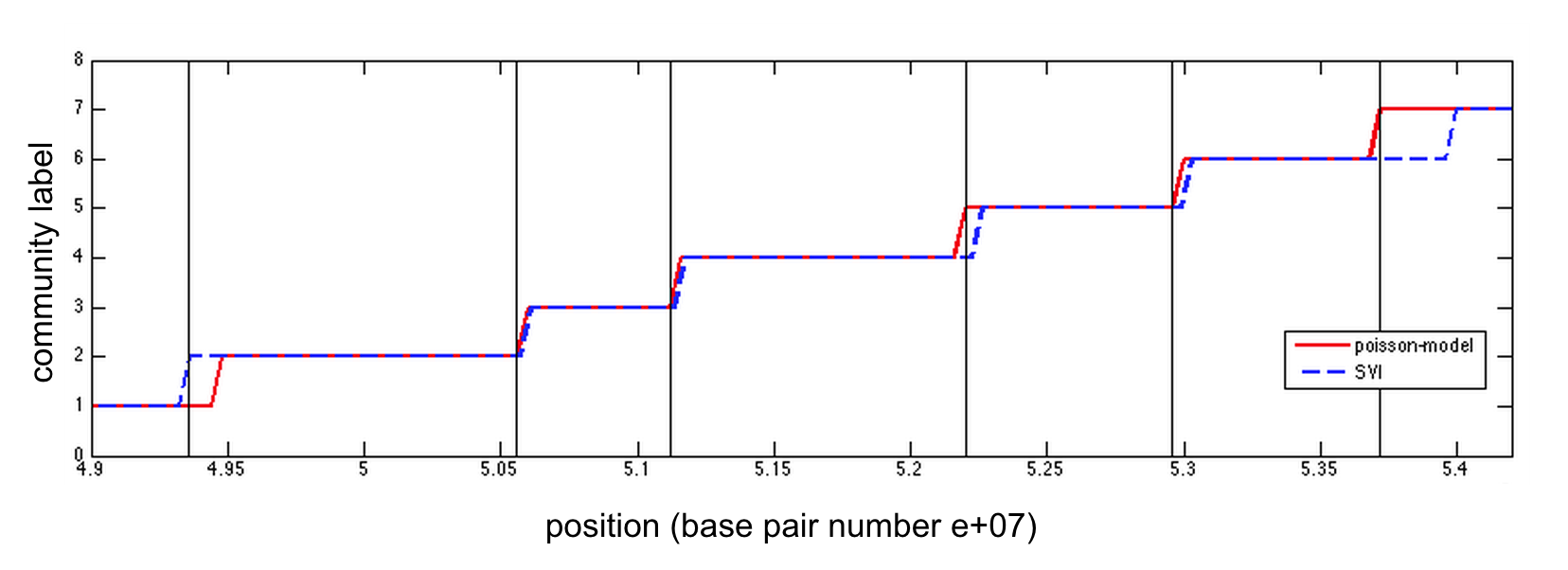} 
   \caption{Outputs of the SVI (dashed blue line) and Poisson (red line) CD algorithms. Though there are always $k!$ equivalent community assignments (corresponding to relabelings of the communities) we require that the community labels rise sequentially along the 1D sequence for ease of comparison between the two algorithms. Black vertical lines correspond to the boundaries identified by Dixon et al. Both algorithms are applied to a portion of the MESC genome presented in the paper by Dixon et al. With $k$ set to 7, both algorithms detect communities that align well with the boundaries discovered by the DI approach. While the boundary between the 6th and 7th communities do not agree between the two CD algorithms, a closer look in Fig. \ref{tads2} shows that both boundary locations correspond to areas of sharp change in the directionality index.}
   \label{tads}
\end{figure}

Fig. \ref{tads} shows that both CD algorithms discover similar (and in some cases identical) TAD boundaries to those identified by the directionality index method. The data is from chromosome 6 of a mouse embryonic stem cell (MESC). In this case we are looking at a near-diagonal region of the chromosome in which there are very few entries of the raw data matrix $X_{ij}$ that are non-zero. Because of this, we had to increase our threshold before applying CD algorithms: specifically, $X^*_{ij} = \mathbbm{1}(X_{ij} > 10)$. We also note that it would be impractical to run CD algorithms on entire genomes with $k\sim10000$ to detect all TADs at once (setting $k$ very high often results in empty communities). In practice, one could search for TADs by running CD algorithms successively on small blocks of the DNA sequence and choose $k$ via the methods suggested in \ref{new} within each new block. 

\begin{figure}[h] 
   \centering
   \includegraphics[width=5in]{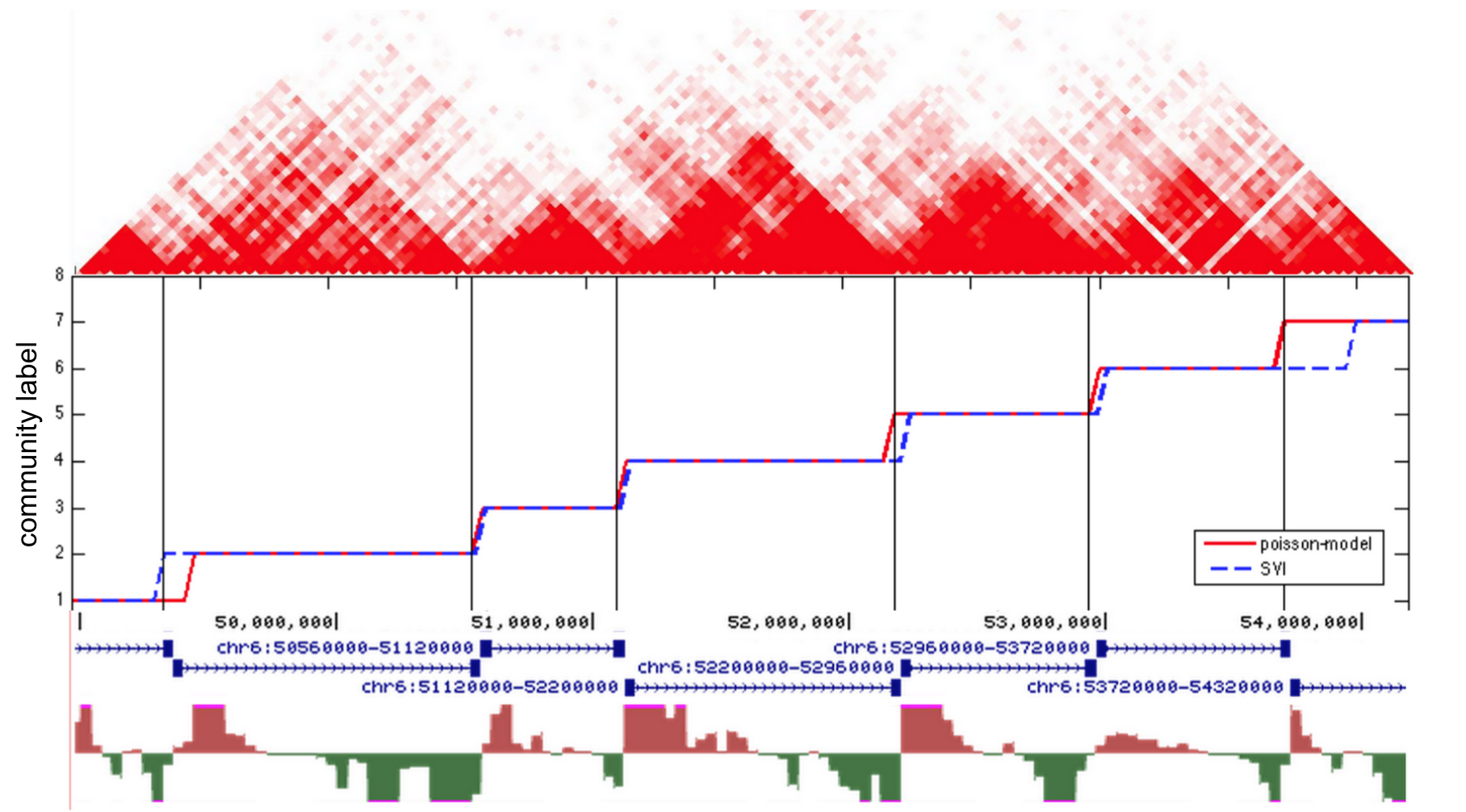} 
   \caption{Comparison between CD outputs, DI output, and original contact matrix. (Top) Original contact matrix of a segment of chromosome 6 from MESC. Only half of the symmetric contact matrix is displayed and it is oriented so that its diagonal lies flat for ease of comparison. (Middle) Identical to \ref{tads}. We see that the black vertical lines (the boundaries identified by the DI method) as well as the jumps in the community assignments from the CD algorithms align well with triangular sections of the contact matrix that correspond to regions with a high level of interconnectivity (TADs). (Bottom) The output of the DI method from Dixon et al. This method identifies boundaries with jumps in the DI: a metric which is positive/negative for loci that have more connections downstream/upstream of the 1D sequence.}
   \label{tads2}
\end{figure}

\subsection{Determining $k^*$ and discovering new communities}
\label{new}
A common problem of many community detection strategies is that the user must supply the algorithm with the number of communities $k$. While this is a notoriously difficult model selection problem in general, it is particularly subtle for this application of CD in Hi-C data. Here, it is clear that there is not a single ``best" $k$; there is biologically important information about communities on various different scales. For instance if we were to set $k=2$, Lieberman-Aiden et al. \cite{erez} suggest that we would find information about the compartment structure of the entire genome. However, this community assignment would give us little information about the TAD structure of the genome. As we have shown above, CD techniques are flexible enough to capture communities at both of these scales.\footnote{Although exploration of Hi-C data at different scales of genome organization has also been proposed before \cite{filippova}, this computational approach can only identify communities continuous in the linear representation of the genome as in the methods of Dixon, et al.} 

So far we have used our previous knowledge from existing research in order to set $k$. In this section, we propose two methods for suggesting meaningful values for $k$ to input into CD algorithms in order to uncover community structures at new scales. The first approach makes use of comparing over many values of $k$ the log-likelihood of a real Hi-C network to that of a random network. The second approach is a more combinatoric approach that aims to quantify the level of redundancy in community detection outputs with $k$ and $k+1$ communities. 

\subsubsection{Approach 1: asymptotic behavior of log-likelihood for large $k$}
\label{A1}
\begin{figure}[!tbp]
  \centering
  \begin{minipage}[b]{0.49\textwidth}
    \includegraphics[width=\textwidth]{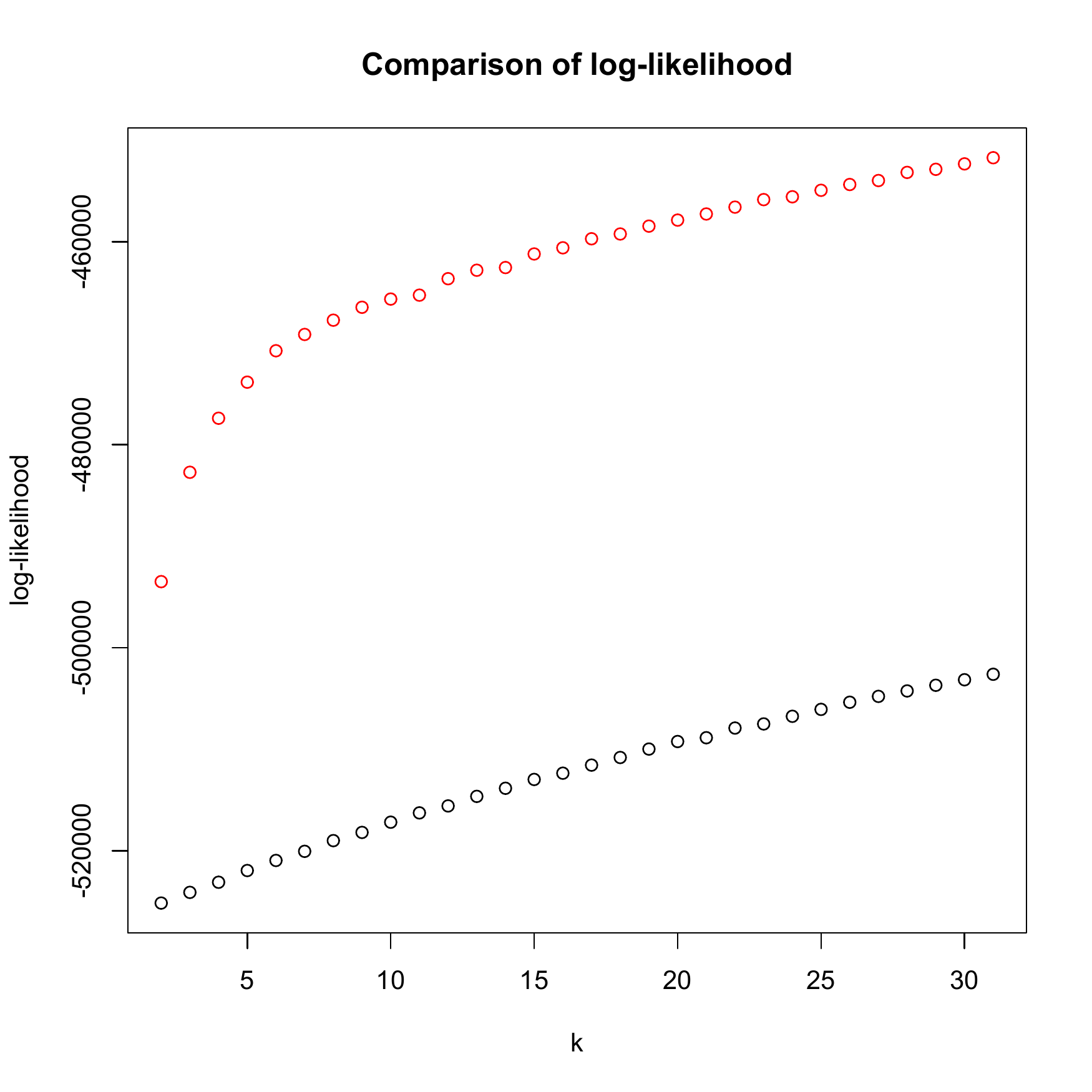}
    \caption{Log-likelihood as a function of $k$ for Hi-C data (red) and a random graphs (black) with an equal number of nodes and an expected number of edges equal to the number of edges in the contact matrix. We see that the log-likelihood of a random graph increases roughly linearly and that for large $k$, the log-likelihood of the data also appears to increase linearly with an equal slope.}
    \label{ll_loops1}
  \end{minipage}
  \hfill
  \begin{minipage}[b]{0.49\textwidth}
    \includegraphics[width=\textwidth]{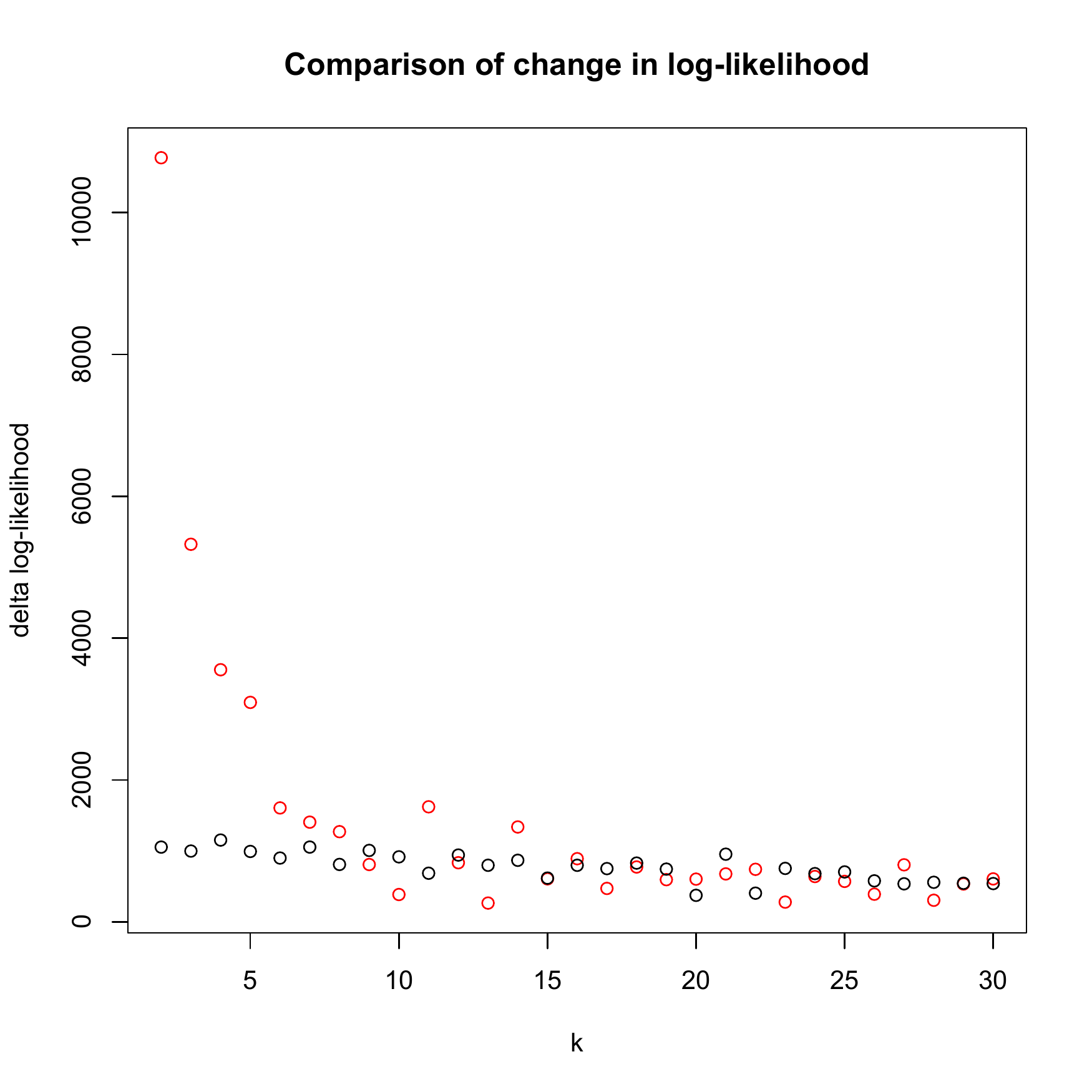}
    \caption{Change in log-likelihood for Hi-C data (red) and a random graph (black) between $k$ and $k+1$ communities. Here we see that the increases in log-likelihood for the random graph are roughly constant and that the increases in the log-likelihood of the data approach this constant value for large $k$.\\
    ~ \\
    ~}
    \label{ll_loops2}
  \end{minipage}
\end{figure}

As we increase $k$, we will inevitably see a consequent increase in the maximum likelihood of the data. To see this, suppose we have found the maximum likelihood assignments of loci for $k$ communities. An assignment to $k+1$ communities can be made with equal likelihood by simply assigning zero members to community $k+1$ and retaining the previous assignments for all other communities. In order to assess whether or not an increase in likelihood is important, we ask whether or not it is distinguishable from the increase we would detect in an ``equivalent" random graph. We define a random graph $A$ as ``equivalent" to another graph $B$ if it has the same number of nodes and the same number of expected edges. Specifically, if $B$ has $n$ nodes and $m$ edges, then $A$ is ``equivalent" if it is a random graph drawn from $G(n,p)$ with $p=\frac{m}{\binom{n}{2}}$. Since there are $\binom{n}{2}$ possible edges in a graph with $n$ nodes, we see that the expected number of edges of such a graph is $p\binom{n}{2} = m$, as required. Fig. \ref{ll_loops1} displays the log-likelihood as a function of $k$ for both chromosome 14 data from Lieberman-Aiden et al. (in red) and an equivalent random graph (in black). Fig. \ref{ll_loops2} suggest that the increase in the log-likelihood for large $k$ is identical to that of an equivalent random graph. It appears that after $k=6$ or $k=7$, the increases in log-likelihood are marginal. Therefore Approach 1 suggests a choice of $k$ somewhere in the range of 6-9. We will refine this choice using Approach 2 (described below) to $k=6$.

\subsubsection{Approach 2}
\label{A2}
\begin{figure}[h] 
   \centering
   \includegraphics[width=4in]{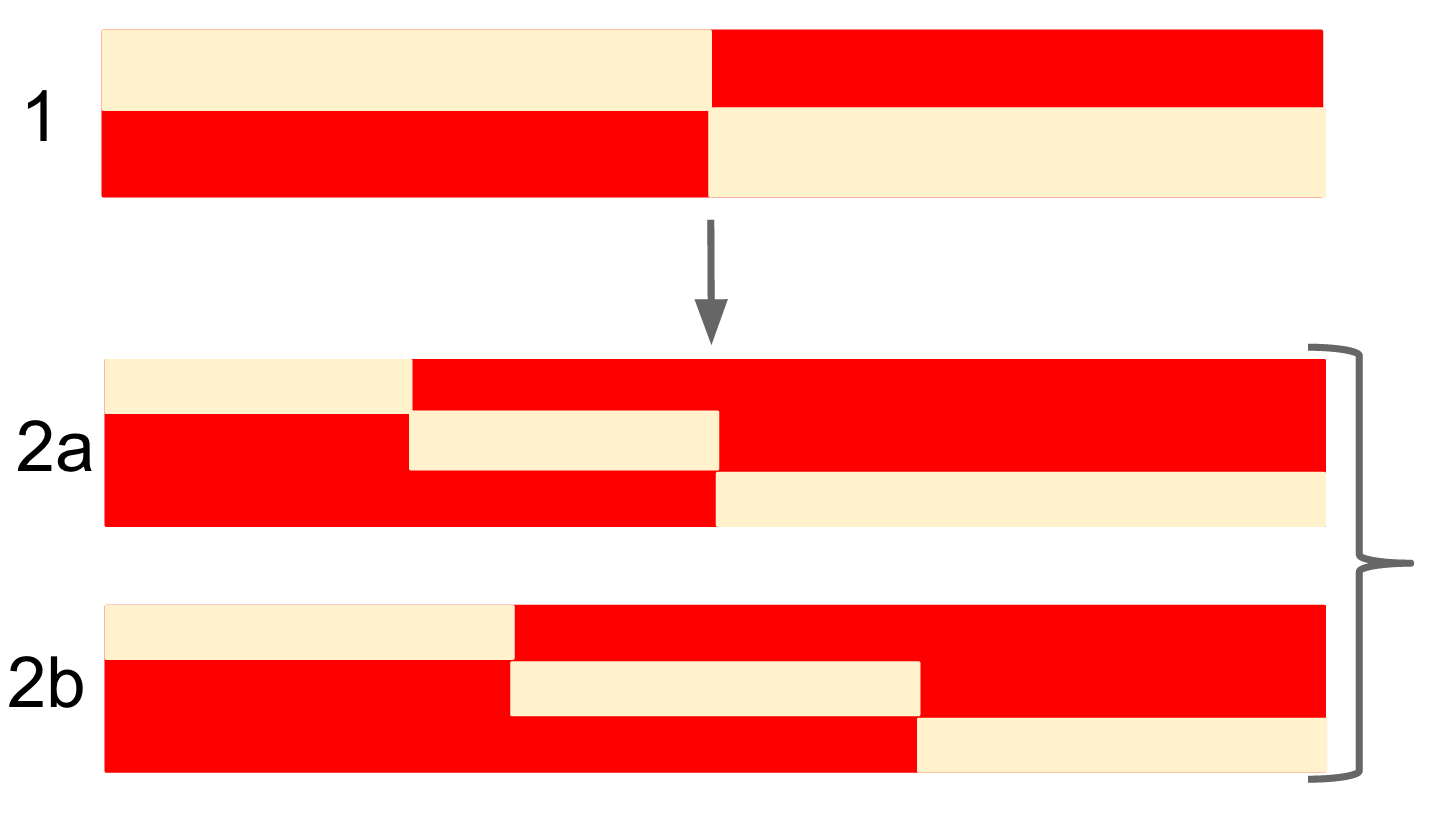} 
   \caption{Toy data for possible transitions between $k=2$ and $k=3$ groups. Each community is a row of this matrix and the columns index genome position. Beige indicates community assignment to that row. A transition from 1 to 2a is evidence for a stable community assignment in 1 whereas a transition from 1 to 2b is evidence for an ``unstable" community assignment in 1.}
   \label{toy_data}
\end{figure}

In order to motivate Approach 2, it will be useful to first consider the toy data displayed in Fig. \ref{toy_data}. 1 represents the community assignment matrix discovered when $k=2$ (where, as before, we assign each locus to the community with the highest admixture proportion). Community assignment 1 discovers two equal-sized communities: one on the left half of the genome and the other on the right half of the genome. 2a and 2b represent two possible community assignment matrices when we increase $k$ to 3. In community assignment 2a the original left community is split into two further sub-communities. In community assignment 2b, three communities are found with the middle community composed of DNA from both the original left and right communities. 

We consider the transition from 1 to 2a as evidence that the two original communities were ``stable." The left community in 1 was split into two communities in 2a either arbitrarily in order to satisfy $k=3$ or because the left community had two sub-communities. Conversely, we consider the transition from 1 to 2b as evidence that the original two communities were ``unstable." The partition into left and right communities in 1 may have appeared arbitrarily in order to satisfy the condition $k=2$. However, when we increase $k$ to 3, we see that there were in fact three distinct communities. We define a metric called the ``assignment instability" which we shall denote $D(k,k+1)$ that aims to quantify this behavior. $D(k,k+1)$ takes as inputs the community assignment matrices of for $k$ and $k+1$ and outputs a measure of the ``stability" of communities discovered for $k$. A low value of $D(k,k+1)$ implies that the community assignments for $k$ communities were stable while a high value indicates unstable assignments for $k$ communities. We describe this metric and its implementation in appendix \ref{hamming}. We emphasize that $D(k,k+1)$ tells us only about the comparative stability between community assignments $A_k$ and $A_{k+1}$. One could imagine generalizing this metric to be of the form $D(k,k+l)$ to describe the assignment stability between community assignments for $k$ and $k+l$ communities. While this may be a useful thing to consider, we do not do this here since we intend to use the assignment instability metric to ``fine-tune" the suggested $k$ from Approach 1. Our strategy will be to use Approach 1 to scan through many values of $k$ and suggest several appear at the ``pivot" of the log-likelihood plot, and then apply Approach 2 to distinguish between these candidates. 

\begin{figure}[!tbp]
  \centering
  \begin{minipage}[b]{0.49\textwidth}
    \includegraphics[width=\textwidth]{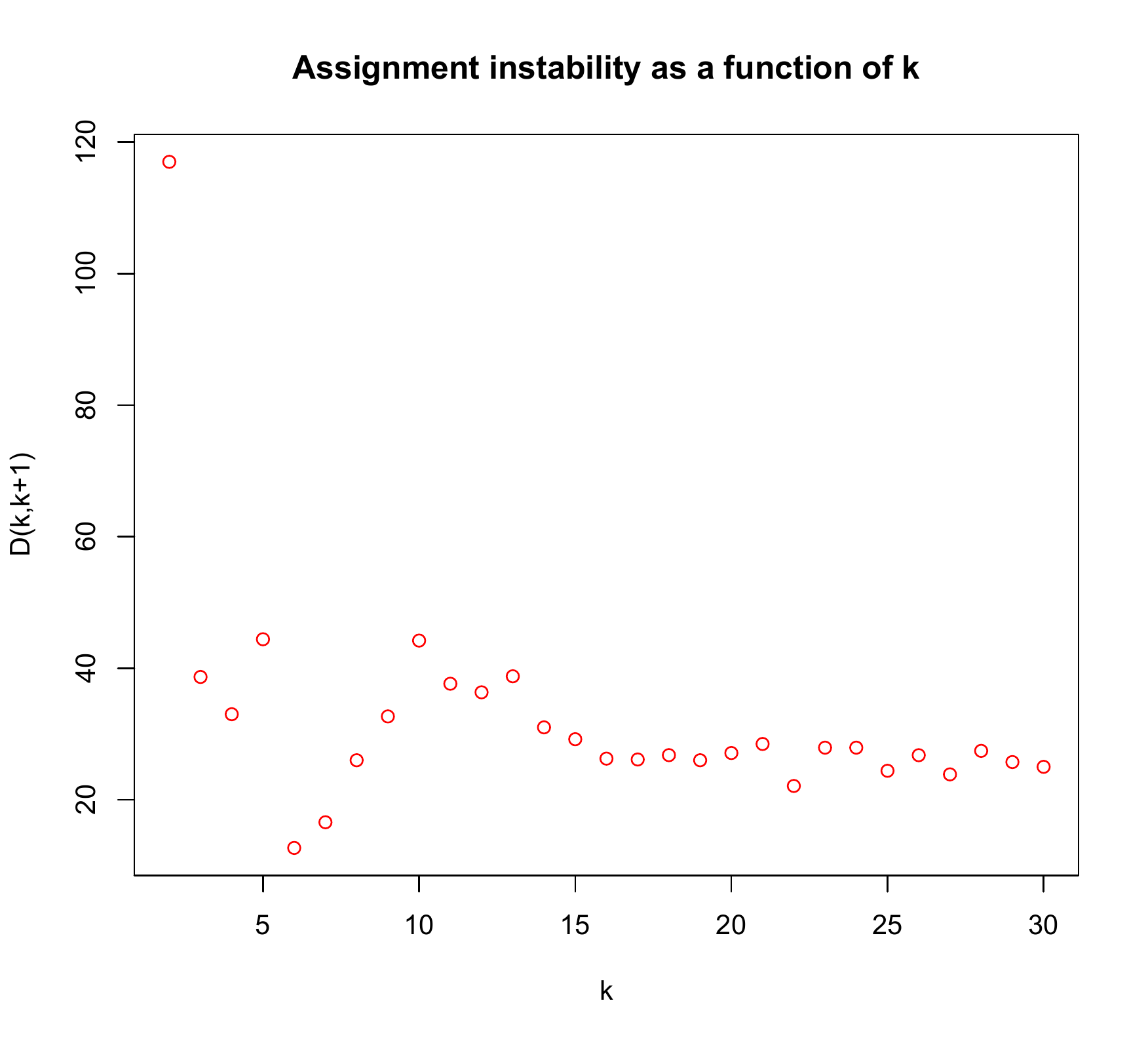}
    \caption{The function $D(k,k+1)$ plotted for human chromosome 14. Large values indicate unstable community assignments while small values indicate stability. The global minimum occurs at $k=6$ which further supports our choice of $k^*=6$. We also note that $k=22$ has a larger assignment instability than $k^*=6$ and $k=7$ only, both of which exhibit similar communities.  Therefore, $k=22$ may be another interesting scale at which meaningful communities emerge and the corresponding community assignment matrix is shown in Fig. \ref{k22}. \\
    ~}
    \label{hamming1}
  \end{minipage}
  \hfill
  \begin{minipage}[b]{0.49\textwidth}
    \includegraphics[width=\textwidth]{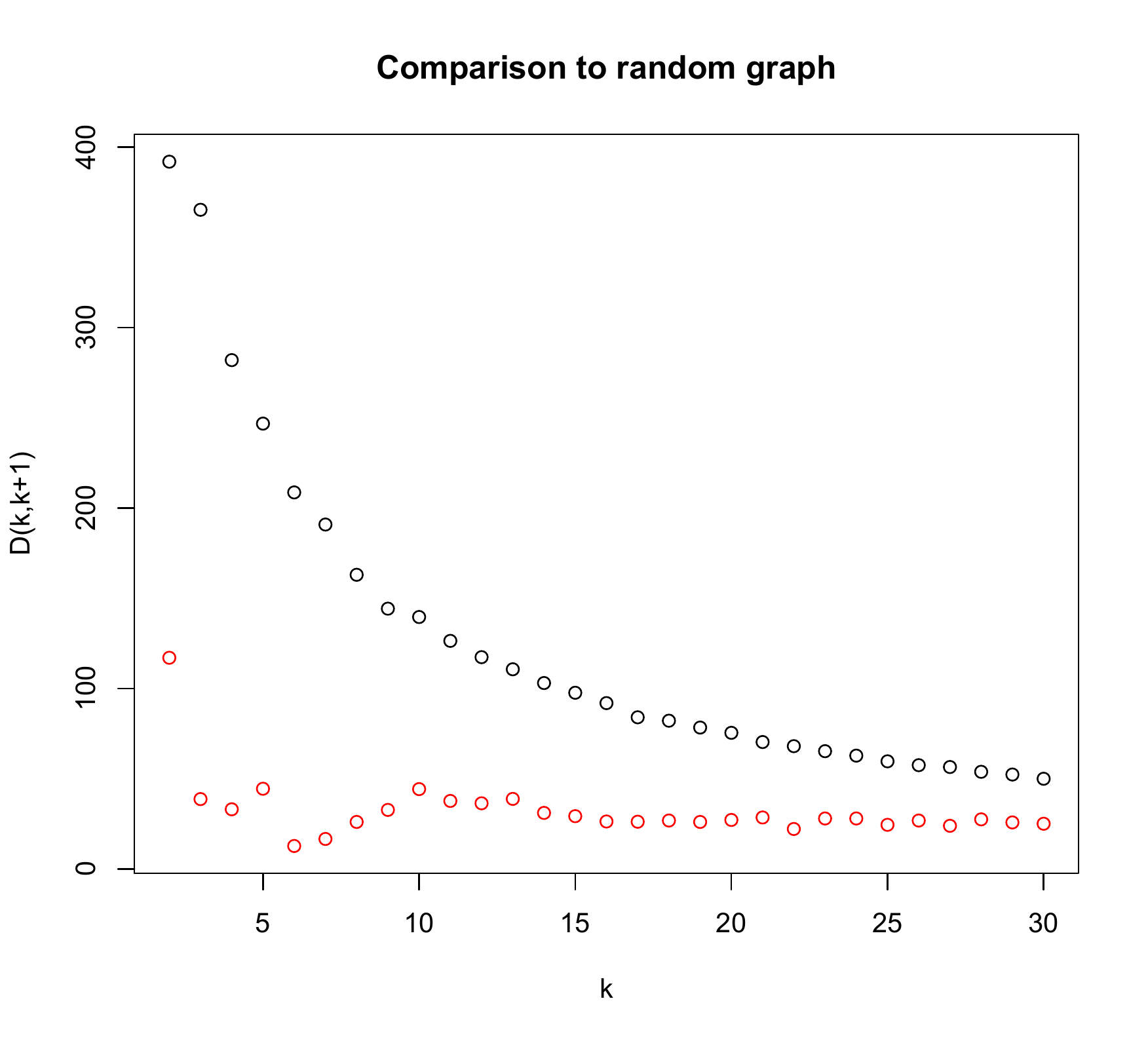}
    \caption{The function $D(k,k+1)$ plotted for human chromosome 14 and a random graph with the same number of nodes and expected number of edges equal to the observed number of edges. We see that the community assignments are uniformly more stable for the real data than the random graph as we would expect since any community structure detected in the random graph occurs by chance. Furthermore, we see that the two function appear to converge for large $k$.\\
    ~\\
    ~
    }
    \label{hamming2}
  \end{minipage}
\end{figure}

Fig. \ref{hamming1} displays the $D(k,k+1)$ and Fig. \ref{hamming2} shows this same plot with the result for a random graph for comparison. The lowest value is found for $k=6$, suggesting high stability in the communities found for $k=6$, consistent with our previous finding using Approach 1. It is interesting to note that if we had used Approach 1 alone, the choice between $k=6$ and $k=7$, for example, would have been arbitrary. In conjunction with Approach 2 however, it is clear that $k=6$ is a superior choice as it has the lowest assignment instability of any $k$. In Fig. \ref{k2k6}, we plot the community assignment matrices for $k=2$ alongside $k=6$. We see that the partition of chromosome 14 found for each $k$ display both stable groups in the $k=2$ assignment (i.e. transitions of the type 1 $\to$ 2a) as well as unstable groups (i.e. transitions of the type 1 $\to$ 2b); for $k=6$, CD finds novel communities that are not simply sub-communities of the original $k=2$ communities discovered by Lieberman-Aiden et al. Approach 2 also highlights other potential $k$ of interest that are not immediately obvious by utilizing Approach 1 alone. For instance, the sharp dip at $k=22$ suggest that this may also be a natural choice of scale. The community assignment matrix for this choice is shown in supplementary Fig. \ref{k22}.

\section{Conclusions and Future Directions}
Two CD algorithms are applied to Hi-C data and compared with the results from two separate studies. We find that both algorithms converge to similar (or identical) outputs for community membership and agree with the independent methods of each study. When we require a partition into two communities, we reproduce the ``compartment" structures found by Lieberman-Aiden et al. However, when we set $k$ to $k^*=6$, we discover new communities. We see that the methods of community detection enable the identification of a variable number of communities as well as potentially non-local communities: features that were not simultaneously possible with either of the pre-existing methods examined here. We further suggest two approaches that leverage our ability to find varying numbers of communities through community detection to suggest natural values for $k$. We note that one particular choice for $k$ that is suggested by both methods is $k=6$ for chromosome 14 of the Lieberman-Aiden et al. data.

Though we have utilized CD algorithms that fit admixture models throughout this work, we note that we do not take full advantage of these models and instead make hard assignments for each locus to an individual community. This capability of mixed-membership models may be useful in future studies in which one seeks individual loci that interact with several different communities. It is plausible that identifying such loci will prove useful for identifying candidate loci that undergo structural repositioning in time. A large variety of block models have been proposed in the literature \cite{airoldi,label_marc,jiaming,newman2,abbs}, where non pairwise interactions, censored measurements, or edge labels are allowed. It would thus be interesting to further investigate and compare these for community detection in Hi-C data sets. 

\section*{Acknowledgement}
The second author would like to thank Aurelie Lozano for initiating discussions on the use of community detection methods in Hi-C data, for providing access to a preliminary set of Hi-C data, and for stimulating discussions and comments on this manuscript. We would also like to thank Prem Gopalan for stimulating discussions and for providing the implementation of the MMSB algorithm used in this work, as well as Harris Lazaris for discussions on Hi-C data.

\appendix
\label{appendix}

\section{Description of algorithms for detecting overlapping communities}
\label{algos}
\subsection{Algorithm 1: Stochastic variational inference}
\label{MMSB}
\begin{figure}[!tbp]
  \centering
  \begin{minipage}[b]{0.49\textwidth}
    \includegraphics[width=\textwidth]{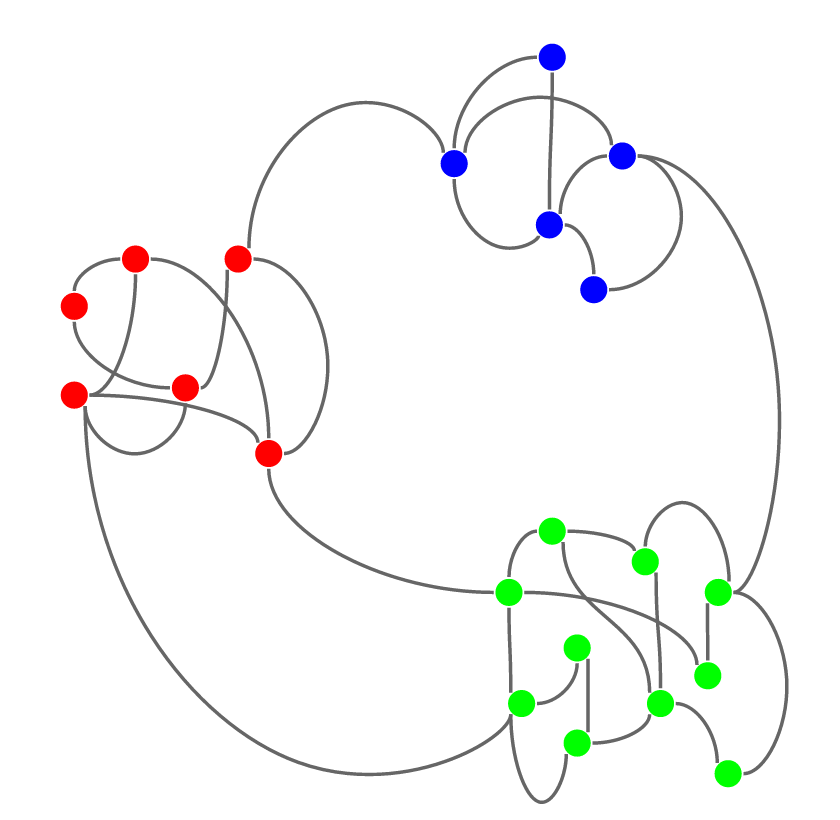}
    \caption{Output for most conventional clustering algorithms. Each node is assigned to a single community (i.e. a color).\\
    ~}
    \label{toynet1}
  \end{minipage}
  \hfill
  \begin{minipage}[b]{0.49\textwidth}
    \includegraphics[width=\textwidth]{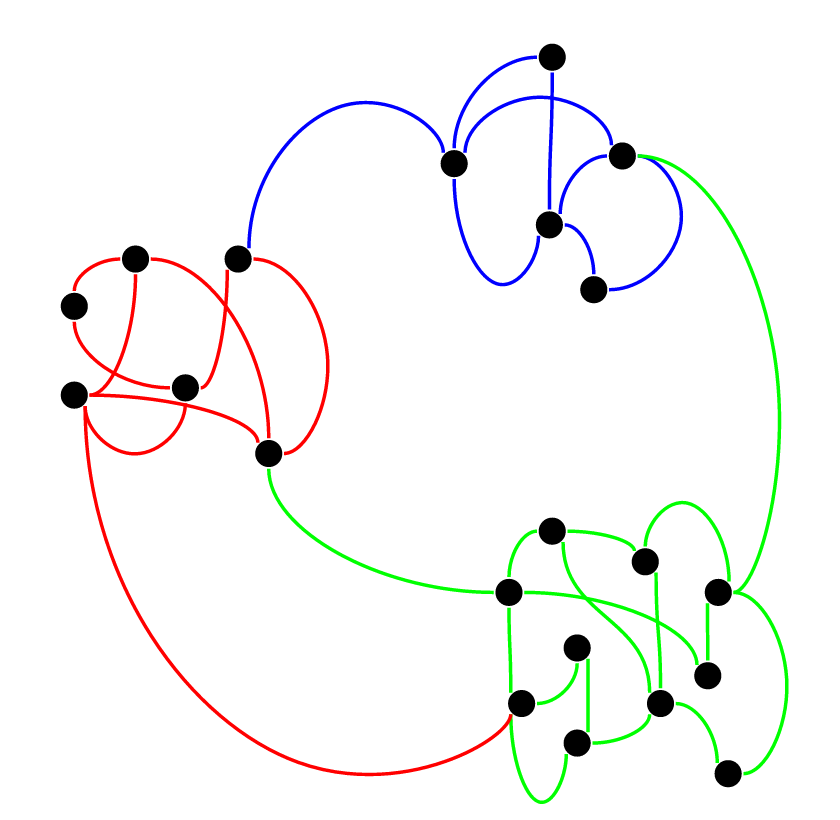}
    \caption{Output for both the SVI and Poisson algorithms. Each edge is assigned to a community. Individual nodes are capable of having an edge from more than a single community.}
    \label{toynet2}
  \end{minipage}
\end{figure}

\paragraph{The model:}
The conventional stochastic block model assigns each node to one of $K$ communities. In the mixed-membership stochastic block model (MMSB) employed by this algorithm, each node $i$ has associated with it a discreet probability distribution $\theta_i$ over the $k$ nodes. For each pair of nodes $(i,j)$, a random variable $z_{i\to j}$ is drawn from the distribution $\theta_i$ and a random variable $z_{j\to i}$ is drawn from $\theta_j$. If $z_{i\to j} = z_{j\to i} = k$, then a link is made between $(i,j)$ with probability $\beta_k$. $\beta_k$ represents the connectivity of group $k$. If $z_{i\to j} \ne z_{j\to i} = k$, then a link is made with a small probability $\epsilon$. Therefore, we have:

\begin{align*}
p(X_{ij}=1|z_{i\to j}, z_{j\to i}) = \left \{ \begin{array}{ll}
\beta_{z_{i\to j}} &: z_{i\to j} = z_{j\to i} \\
\epsilon &: z_{i\to j} \ne z_{j\to i} \\
\end{array}
\right.
\end{align*} 
\noindent
where $X$ is the adjacency matrix. In this model, as in the Poisson model described below, the edges rather than the nodes belong to a single group. The fundamental difference between both of these models and conventional clustering algorithms is made clear in Fig. \ref{toynet1} and \ref{toynet2}. In both of the algorithms considered here, a single node (i.e. a DNA locus) is capable of having an edge from more than a single community, however in this work we do not exploit this flexibility and instead assign each locus to the community it shares the largest number of edges with. 

\paragraph{The algorithm:}
The task is now to compute the posterior probability $p(\bf{\theta},\bf{z}|\bf{X}) = p(\bf{\theta}, \bf{z}, \bf{X})/p(\bf{X})$ and find the set of parameters which maximize this posterior. This is a difficult problem however because of the intractable denominator

\begin{align*}
p(\bf{y}) = \int_{\bf{\theta}}\sum_{\bf{z}}p(\bf{\theta},\bf{z},\bf{X})
\end{align*}

\noindent

Rather than computing this posterior probability distribution directly, the authors in \cite{prem} recast this problem into a mean-field variational inference setting. They show that minimizing the KL distance between the true posterior is equivalent to maximizing the objective function

\begin{align*}
\mathcal{L}(\bf{\gamma},\bf{\phi}) = \E[\log p(\bf{\theta},\bf{z},\bf{X})]+H[q(\bf{\theta},\bf{z})]
\end{align*}

\noindent
where $q$ is a member of a variational family of posterior distributions and expectation occurs over this function. Maximizing this function rather than maximizing the original posterior probability directly has the advantage that it can be done using several pre-existing optimization methods. The method that the authors use is that of stochastic optimization, in which a local minimum is found by following a noisy approximation of the gradient. A full description of the algorithm may be found in \cite{prem} and a downloadable implementation is available at: \url{https://github.com/premgopalan/svinet}. 

\subsection{Algorithm 2: Poisson model}
\label{poisson}
\paragraph{The model:}
The model used in this algorithm is similar in spirit that used by the SVI algorithm. As with the SVI model, this model assigns a unique group to each edge rather than each node and each node $i$ comes with a probability distribution $\theta_i$ over the $K$ communities. Unlike the model used by the SVI algorithm, this model allows multiple edges from the same community to link two nodes so that each pair of nodes $(i,j)$ has a mean $\theta_i(k)\theta_j(k)$ links via community $k$. The number of links between two nodes from a single community can be approximated by a $\text{Pois}(\theta_i(k)\theta_j(k))$ distribution. In this current implementation of the Poisson model, all adjacency matrices are treated as binary: $X_{ij}=1$ for all contact matrix entries that are non-zero. This has been done in part for simplicity of comparison between the Poisson model and the MMSB model.

\paragraph{The algorithm:}
Given the above model, one can write down the probability of a graph $G$ with adjacency matrix $\bf{X}$ and community membership distributions $\{\theta_i\}$ as 

\begin{align*}
P(G|\theta) = \prod_{i<j}\frac{\left(\sum_z \theta_i(z)\theta_j(z)\right)^{X_{ij}}}{X_{ij}!}\exp\left(-\sum_z \theta_i(z)\theta_j(z)\right) \times \prod_{i}\frac{\frac{1}{2}\left(\sum_z \theta_i(z)^2\right)^{X_{ij}}}{(X_{ij}/2)!}\exp\left(-\frac{1}{2}\sum_z \theta_i(z)^2\right)
\end{align*}

\noindent 
Now the task is maximize over all possible distributions of $\{\theta_i\}$ defined by $\sim n \times k$ parameters. The authors then show that solving this optimization problem is equivalent to solving the following system of equations

\begin{equation*}
\begin{array}{lcl}
q_{ij}(z) &=& \frac{\theta_i(z)\theta_j(z)}{\sum_z\theta_i(z)\theta_j(z)} \\ 
\theta_i(z) &=& \frac{\sum_j A_{ij}q_{ij}(z)}{\sum_i \theta_i(z)}
\end{array}
\end{equation*}

\noindent
which are equivalent to the ``E" and ``M" steps of the EM algorithm. This system of equations can be solved by starting with arbitrary values for the distributions $\{\theta_i\}$ and alternating between the two equations until convergence. This method increases the likelihood of $P(G|\theta)$ with each iteration, although it is not guaranteed to converge on the global maximum. A full description of this algorithm may be found in \cite{ball} 

\section{Description of $D(k,k+1)$}
\label{hamming}
We assume that we have the community assignment matrices for $k$ and $k+1$. We consider all $\binom{k+1}{2}$ ways in which to combine 2 of the $k+1$ rows by defining a $(k+1)\times k$ matrix $B$ such that the first column is a particular 0-1 vector with two ones (the index of these ones corresponds to the rows to be combined) and the remaining columns have a single one in the remaining $k-1$ rows that do not have a 1 in the first column. For example, if we are combining rows 3 and 5 and $k=5$, one choice for $B_{(3,5)}$ is:

\begin{align*}
B_{(3,5)} = \begin{pmatrix}
0 & 1 & 0 & 0 & 0\\
0 & 0 & 1 & 0 & 0\\
1 & 0 & 0 & 0 & 0\\
0 & 0 & 0 & 1 & 0\\
1 & 0 & 0 & 0 & 0\\
0 & 0 & 0 & 0 & 1\\
\end{pmatrix}.
\end{align*}

Denoting $A_{k+1}$ for the $k+1$ community assignment matrix, we then have for each combination of rows $(a,b)$ the matrix $\tilde{A}_k = B_{(a,b)}^T A_{k+1}$. We can then compare the Hamming distance of each row of $\tilde{A}_k$ to each row of $A_k$. These distances form a $k\times k$ matrix $H$.We next note that the model is non-identifiable in the sense that there are $k!$ equivalent outputs that we could have received from the CD algorithms corresponding to different labelings of communities. Finding the best alignment is then equivalent to finding the minimal sum of entries from $H$ such that a single entry is chosen from each row and column. This is identical to assignment problem and is solved by the Hungarian Method. Finally, we choose the combination with the minimum value over all $\binom{k+1}{2}$ choices. Finally, we normalize this value by dividing by $k$. In doing this, we characterize the average level of instability of each group. This final value is defined as $D(k,k+1)$. 

\section{Supplementary Figures}
\label{sup_figs}
\begin{figure}[h] 
   \centering
   \includegraphics[width=3in]{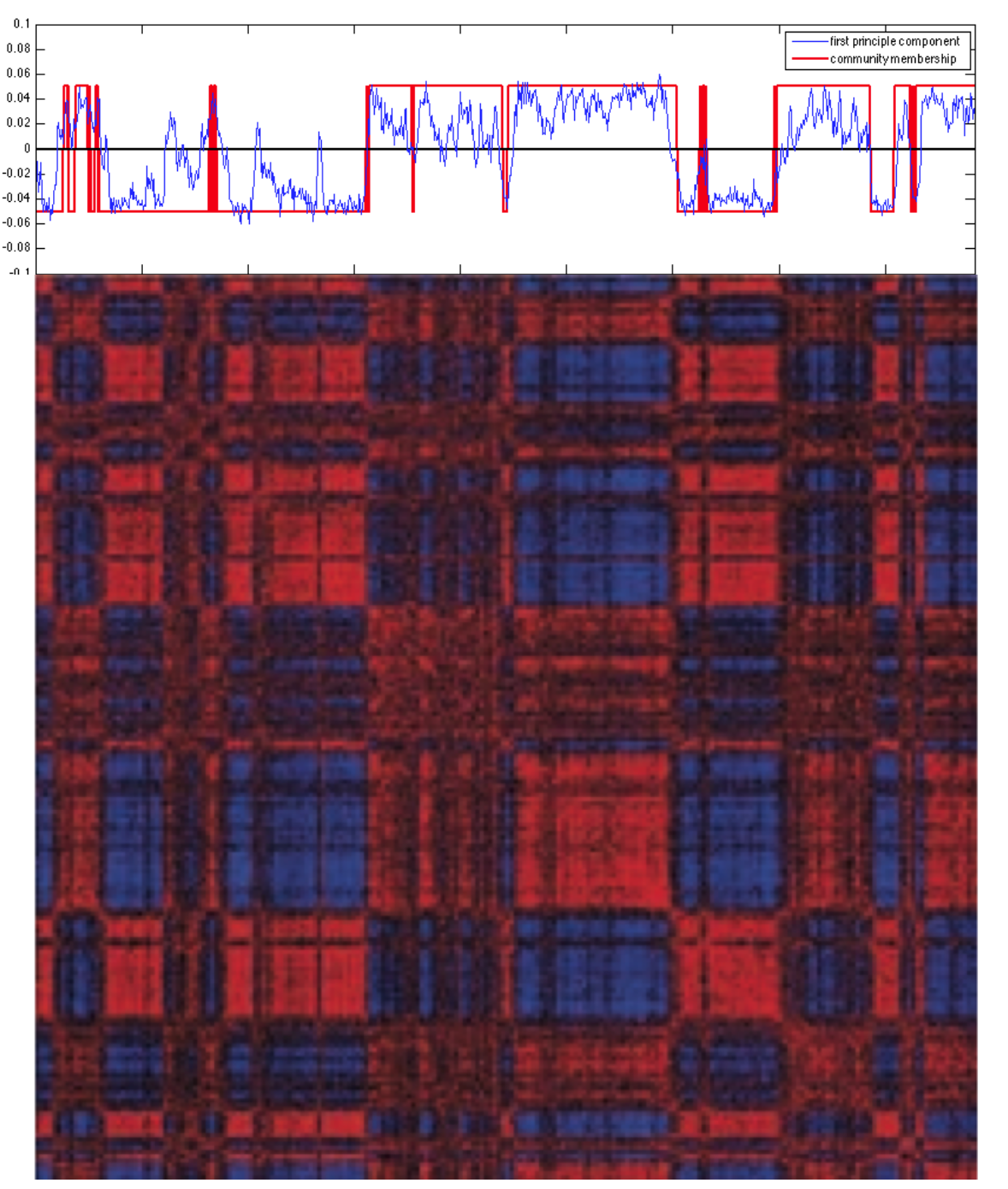} 
   \caption{Comparison between Pearson correlation of Hi-C contact matrix and results of PCA and community detection methods. We see that the community boundaries align closely to the boundaries of the ``plaid" pattern of the Pearson corralation matrix.}
   \label{princomp2}
\end{figure}

\begin{figure}[h] 
   \centering
   \includegraphics[width=4in]{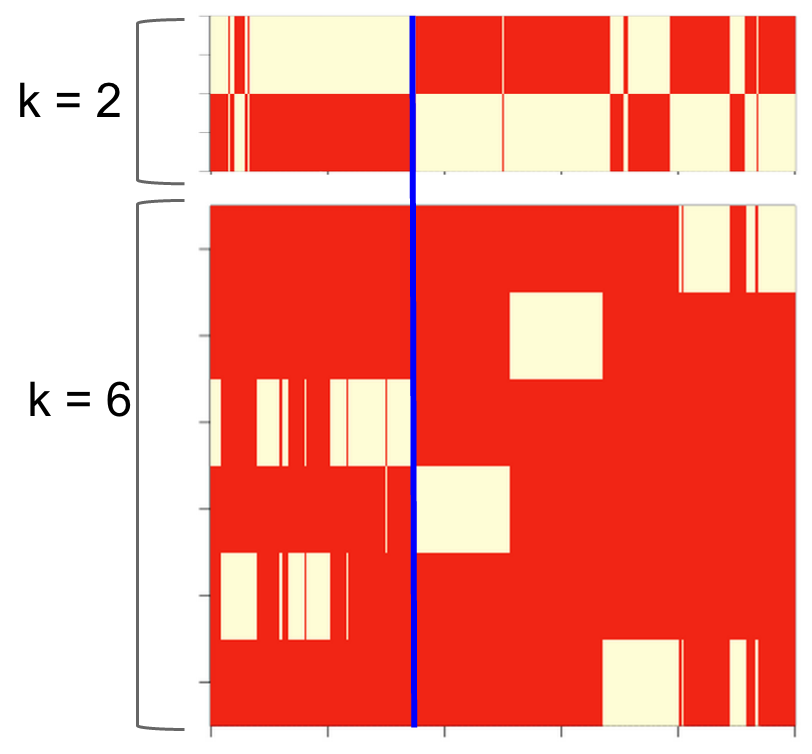} 
   \caption{Community assignment matrices for $k=2$ (top) and $k=6$ (bottom). To the right of the blue line, it appears that most communities in the $k=6$ matrix share boundaries with those from the $k=2$ matrix (indicating assignment stability). To the left of the blue line, the community boundaries appear to be distinct between the two matrices (indicating assignment instability).}
   \label{k2k6}
\end{figure}

\begin{figure}[h] 
   \centering
   \includegraphics[width=4.5in]{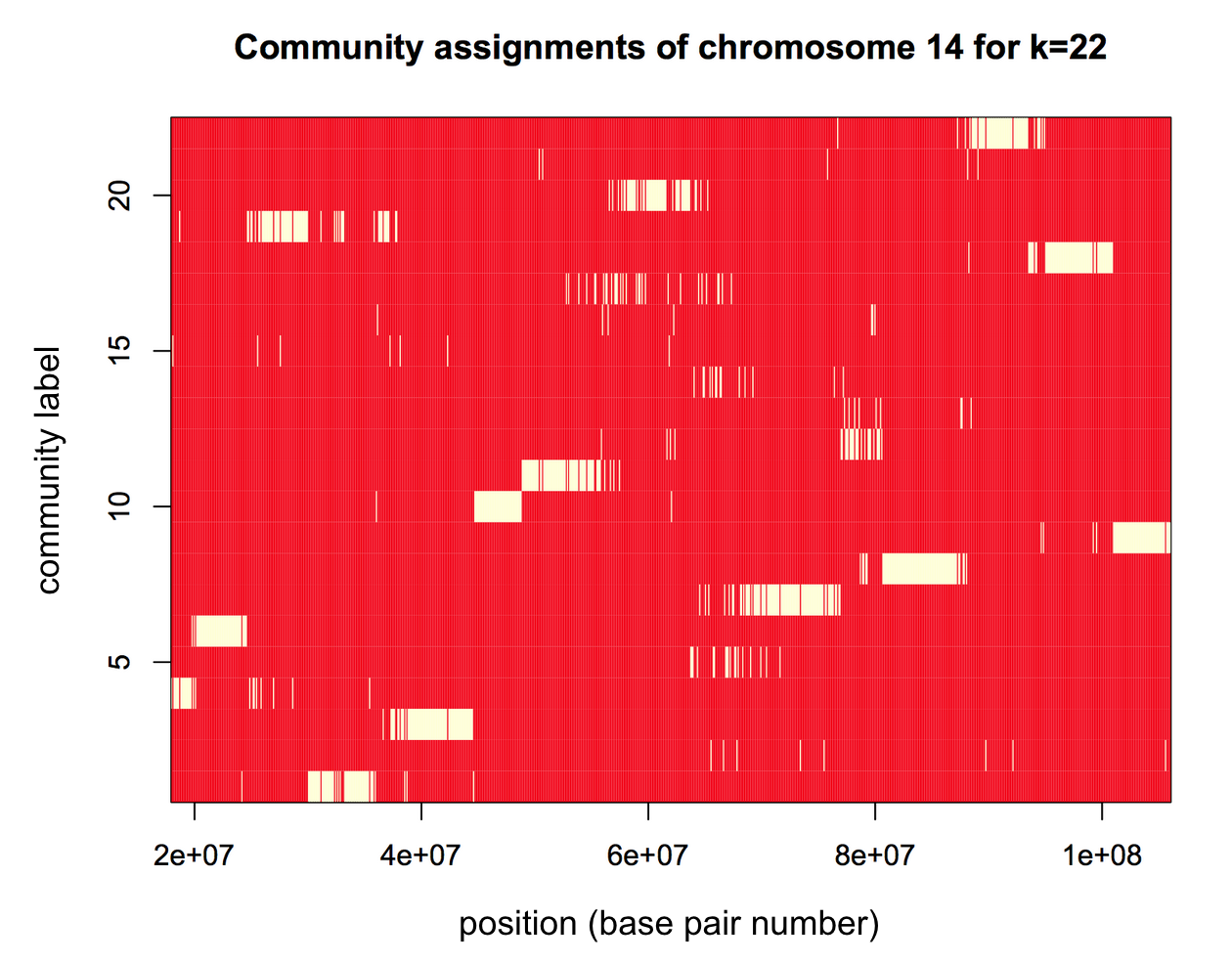} 
   \caption{Community assignment matrix for $k=22$.}
   \label{k22}
\end{figure}

\end{document}